\documentclass[showpacs,preprint]{revtex4}
\usepackage{amsmath}
\usepackage{graphicx}
\usepackage{amsfonts, amssymb}
\usepackage{gensymb}
\usepackage{cool}
\usepackage{float}
\def\exp{\text{e}}
\def\be{\begin{equation}}
\def\ee{\end{equation}}

\def\Hef{\vec{H}_{\mbox{{\tiny eff}}}}

\def\hef{\vec{h}_{\mbox{{\tiny eff}}}}

\def\kpar{k_u}
\def\kperp{k_d}
\begin{document}

\title{ Reaction diffusion dynamics and the Schryer-Walker solution for domain walls of the Landau-Lifshitz-Gilbert equation }

\author{R. D. Benguria}
\author{  M. C. Depassier}
\affiliation{Instituto de F\'\i sica, Pontificia Universidad Cat\'olica de Chile \\ Casilla 306, Santiago 22, Chile}

%{Compiled on {\ddmmyyyydate\today} at \currenttime}
% \date{   \currenttime \today}

\begin{abstract}
We study the dynamics of the equation obtained by Schryer and Walker for the motion of domain walls. The reduced equation is a reaction diffusion equation for the  angle  between the applied field and the magnetization vector. If the hard  axis  anisotropy $K_d$  is much larger than the easy axis anisotropy $K_u$,  there is a range of applied fields  where the dynamics does not select the  Schryer-Walker solution. We give analytic expressions for the speed of the domain wall in this regime and the conditions for its existence.
\end{abstract}
%\pacs{75.78.Fg}{Dynamics of domain structures}
%\pacs{75.78.-n}{Magnetization dynamics}

\pacs{75.78.-n, 75.78.Fg }

\maketitle

Magnetic domain wall propagation is an active area of research both as an interesting physical phenomenon as well as for its possible applications in logic devices, magnetic memory elements and others \cite{Stamps2014}.       In micromagnetic theory  the motion of a domain wall  is described by the Landau Lifshitz Gilbert (LLG) equation \cite{Landau1935,Gilbert1956} which cannot be solved analytically except in few special cases.    For an infinite medium with uniaxial anisotropy and an external field applied along the symmetry axis, the Schryer Walker   (SW) solution \cite{Walker, ScWa1974}  is the best known analytical expression for  a stationary traveling  domain wall.    This  exact solution  predicts successfully many experimental results  below  a cut-off field.
The stability of the SW solution with respect to small perturbations  has been studied  recently \cite{HuWa2013} using dynamical systems techniques. The analysis of the spectrum of a perturbation to the SW solution shows that it may become absolutely or convectively  unstable before the breakdown field. This instability is found numerically  for sufficiently large hard axis anisotropy and for fields larger than a critical value. While this instability is in qualitative agreement with results of the numerical integrations reported in \cite{WaYa2012}  it has not been confirmed experimentally. The  range of physical parameters of ferromagnetic materials where domain walls are observed  is wide and, as discussed in \cite{HuWa2013},  it has not been fully explored.   Temperature, doping and fabrication techniques allow tailoring of the material parameters which may vary over several orders of magnitude \cite{ThHu2012,  Zhang2014, XuYi2015} and such instability  may become experimentally accesible in the future  \cite{WaYa2012}.

In this work we study the dynamics of the equation derived by Schryer and Walker  from the LLG problem. Most work on this equation has focused on the existence of an exact traveling domain wall.  Here we take a different approach, the SW equation is a nonlinear partial differential equation  and the existence of an exact solution does not imply that an initial condition will converge to this exact solution.  Using the theory of reaction diffusion equations  we find  that   the SW solution  is not selected by the dynamics   for applied fields larger that a a critical value  provided that the hard axis anisotropy is sufficiently large. We give explicit analytic   expressions for the transition field, for the speed of the front beyond this transition and the conditions under which this transition occurs.  Qualitatively these results agree with the findings reported in \cite{HuWa2013};  in this case, due to the simplicity of the problem,  a full analytical solution is given.  We conjecture that this behavior is preserved in the full LLG equation. An asymptotic analysis of the slow time evolution of the   LLG equation for large perpendicular anisotropy and small fields  leads to a similar transition \cite{De2015}.
 
 For the sake of clarity we first recall in some detail the SW solution.
The starting point of the calculation  is the LLG  equation for the magnetization. The material has magnetization $\vec M= M_s \vec m$ where $M_s$ is the saturation magnetization and $\vec m = (m_1,m_2,m_3) $ is a unit vector along the direction of magnetization. 
The dynamic evolution of the magnetization  is governed  by the LLG equation, 
\be\label{LLG}
 \frac{d \vec M }{ d t} = - \gamma_0  \vec M \times \Hef + \alpha\frac{\vec M}{M_s} \times \frac{d \vec M }{ d t}
 \ee

where $\Hef$ is the effective magnetic field,  $\gamma_0=  | \gamma | \mu_0$,   $\gamma$ is the gyromagnetic ratio of the electron and  $\mu_0$ is the magnetic permeability of 
vacuum. The constant  $\alpha>0$ is the dimensionless phenomenological Gilbert damping parameter. Following SW, we consider an infinite medium with uniaxial crystalline anisotropy. The easy axis is taken to be the $z$ axis of a cartesian coordinate system and an external magnetic field  is applied along this easy axis.  The demagnetizing field is assumed to have a local representation and to depend only on $x$. The effective magnetic field is given then by
\be \label{H1}
\Hef = H_a \hat z + \frac{C_{\rm ex}}{\mu_0 M_s^2}  \frac{\partial^2 \vec M}{\partial x^2} + \frac{2 K_u}{\mu_0 M_s^2} M_z\hat z -  \frac{2 K_d}{\mu_0 M_s^2} M_x \hat x,
\ee
where $C_{ex}$ is the exchange constant, $K_u$ the easy axis uniaxial anisotropy and $K_d$ the perpendicular anisotropy.

Introducing $M_s$ as unit of magnetic field, and introducing the dimensionless space  and time variables   $\xi =  x \sqrt{K_u/C_{ex}}$  and $\tau= \mu_0 |\gamma | M_s t$  we rewrite equations (\ref{LLG}) and (\ref{H1}) in dimensionless form

\be\label{LLG1}
 \frac{d \vec m }{ d \tau} = -   \vec m \times \hef + \alpha \vec m  \times \frac{d \vec m }{ d\tau }
 \ee
 with
\be \label{hef}
\hef= h_a \hat z + \frac{1}{2} \kpar  \frac{\partial^2 \vec m}{\partial \xi^2} + \kpar m_3 \hat z - \kperp m_1 \hat x .
\ee
where $h_a$ is the dimensionless applied field and the dimensionless numbers that have appeared are $\kpar= 2 K_u/(\mu_0 M_s^2)$ and  $\kperp= 2 K_d/(\mu_0 M_s^2)$. 
Equations (\ref{LLG1})  and (\ref{hef})  describe the dynamics of the problem. Next we introduce spherical coordinates for the unit magnetization vector, namely, 
\be 
m_1 = \sin\theta \cos\varphi, \quad m_2 = \sin\theta \sin\varphi, \text{  and  } m_3 =\cos\theta.
\ee
The LLG equation then reduces to the coupled system
\begin{subequations}
\begin{align}
 \alpha \sin\theta\,  \dot\varphi + \dot\theta &= \frac{1}{2}\kperp \sin\theta  \sin2 \varphi  + \frac{1}{2}\frac{\kpar}{\sin\theta} \frac{\partial}{\partial \xi} \left(  \varphi_{\xi} \sin^2 \theta\right),
 \\
 \alpha  \dot\theta - \sin\theta \, \dot\varphi  &=  \frac{1}{2} \kpar \theta_{\xi\xi} - h_a \sin\theta   - \frac{1}{4} \kpar  \varphi_{\xi}^2 \sin 2\theta -  \kpar \sin \theta\, \cos\theta  -   \kperp \sin\theta \,\cos\theta\cos^2\varphi.
 \end{align}
 \end{subequations}

The solution studied by Schryer and Walker is that with  constant azimuthal angle $\varphi = \varphi_0.$  With this assumption  the equations above  reduce to
\begin{subequations} \label{SW}
\begin{align}
 \dot\theta & = \frac{1}{2}\kperp \sin\theta   \sin 2 \varphi_0, \label{m1}
 \\
 \alpha  \dot\theta &=     \frac{1}{2} \kpar \theta_{\xi \xi}  - F(\theta)\label{m2} 
  \end{align}
 \end{subequations}
where
\be
 F(\theta) = \sin\theta [  h_a+    \cos \theta  (\kpar +  \kperp \cos^2\varphi_0)]. \label{F}
 \ee

 The time evolution for the polar angle $\theta$ is governed by a reaction diffusion equation, for which a complete rigorous mathematical  theory is well established. We are interested in the reversal of the magnetization induced by the applied magnetic field, therefore, as in \cite{Walker,ScWa1974}, we assume that $\theta_{\xi}$ vanishes as $\xi\rightarrow \pm \infty$ and $\theta \rightarrow 0$ when $\xi\rightarrow -\infty$, $\theta\rightarrow \pi$ when $\xi \rightarrow  \infty$.  Equations  (\ref{SW})  together with the asymptotic conditions are the system studied by Schryer and Walker.  For the sake of completeness we recall some of their results. The first step is to notice  that the  angle $\varphi_0$ is fixed through a consistency condition. In effect, multiplying (\ref{m1}) by $\alpha \theta_x$ and integrating in  $x$ between $-\infty$ and $+\infty$ , using (\ref{m2}) and the boundary conditions, one obtains\cite{Walker}
 \be
 \sin 2 \varphi_0 = -\frac{2 h_a}{ \alpha k_d} \equiv - \frac{h_a}{h_w},
 \ee
 where the dimensionless  Walker field is given by $  h_w = \alpha k_d/2$ in the present notation. 
  It is convenient to express $\cos^2\varphi_0$ in terms of the applied magnetic field. Two branches exist,  
 $
 \cos^2\varphi_0 =   (1/2)  ( 1 \pm \sqrt{1-\sin^2 2\varphi_0}).$  When the applied field vanishes the domain wall is static and  $\varphi_0=\pm  \pi/2$.   Therefore we choose, following \cite{ScWa1974} the branch
$$
 \cos^2\varphi_0 =   \frac{1}{2}  \left( 1 - \sqrt{1-\sin^2 2\varphi_0} \right) =   \frac{1}{2}\left( 1 - \sqrt{1- (h_a/h_w)^2}\right).
 $$

Next we consider the dynamics of equation (\ref{m2}). In order to apply the theory of reaction diffusion equations,  it is convenient to render it in the usual form. To do so  we  introduce a new dependent variable $u$ defined by
 $\theta = \pi(1-u).$ The evolution equation for this new variable is
 \be\label{RDeq}
 \alpha \dot u  =   D u_{\xi \xi}  + f(u), \ee
 with
$$
 f(u) = \frac{\sin\pi u}{\pi} [h_a - (\kpar+\kperp \cos^2\varphi_0)\cos\pi u)],\,\, D= \frac{1}{2} \kpar
$$
 The explicit dependence of the reaction term on the applied field is then
 \be \label{f(u)}
  f(u) = \frac{\sin\pi u}{\pi} \left( h_a -   R(h_a)  \cos\pi u\right), \text{    where  } R(h_a)  = \kpar + \frac{\kperp}{2} \left( 1 - \sqrt{1- (h_a/h_w)^2}\right).
\ee

 Equation (\ref{RDeq}) is the well studied reaction diffusion equation. The diffusion constant $D= \kpar/2$ and the reaction term $f$ which satisfies $f(0) = f(1) =0$ is monostable or bistable depending on the values of the material parameters and the applied field.  In the bistable case, there is a unique domain wall. 
  This is the exact Schryer Walker solution,
  \be \label{SWsol}
   u = \frac{2}{\pi} \arctan\left[ \exp^{ - \sqrt{\frac{2 R(h_a)}{\kpar}}(\xi - c_{\text{\tiny{SW}}} \tau)} \right],
   \qquad  c_{\text{\tiny{SW}}}  = \frac{h_a}{\alpha} \sqrt{ \frac{\kpar}{2 \kpar +\kperp(1 - \sqrt{1-(h_a/h_w)^2})}}.
   \ee

  When the reaction term is monostable ($f'(0)>0$)  there is a continuum of fronts.   A small perturbation to the unstable state $u=0$ ($\theta=\pi$) evolves into  a traveling monotonic front of minimal speed $c^*$ \cite{KPP37,AW78} that joins the  stable state   $u=1 \,  (\theta=0)$ to the  unstable state $u=0 \,(\theta=\pi)$.  The minimal speed can be obtained from a variational principles \cite{HaRo75,BD96} and is bounded  by \cite{AW78}
\be\label{bounds}
c_{\text{\tiny{KPP}}}  \equiv \frac{2}{\alpha}  \sqrt{D f'(0)} \leq c^* \leq \frac{2}{\alpha}  \sqrt{ D \sup f(u)/u }.
 \ee
When the  asymptotic speed is exactly $c_{\text{KPP}} $  the traveling front is called a KPP or pulled front (see \cite{vanSaarloos} for a review).  
 In the monostable case one must determine whether   the front of minimal speed is the  SW solution Eq. (\ref{SWsol}) or a KPP front, of speed
   $$
c_{\text{\tiny{KPP}}}  = \frac{2}{\alpha} \sqrt{ \frac{\kpar}{2}(h_a- R(h_a))}.
   $$
 
 The analysis that follows gives the exact criterion under which the speed of the domain wall will be given by the KPP value. We show below that  for an applied field $H_{-} < H < H_{+}$ where
\be\label{hlimits}
 H_{\pm} = \frac{2 \alpha K_d}{\mu_0 M_s} [ \alpha(1 + 2 \kappa) \pm \sqrt{ \alpha^2 - 16\kappa(\kappa+1)} \, ] \quad  \text{and}\quad \kappa= K_u/K_d.  
 \ee
 the front of minimal speed, which will be selected by the dynamics, is a KPP front. 
 This regime exists  provided that $\alpha>  \sqrt{16 \kappa (\kappa+1)}$.  Otherwise, the speed is given by the Schryer Walker solution. 
 
One can verify that  the reaction term is monostable $f'(0) >0$ for
$$
\frac{ \alpha(1 + 2 \kappa) -\sqrt{\alpha^2 - 4\kappa (\kappa+1)}}{1+ \alpha^2} \le \frac{h_a}{h_w} \le \frac{ \alpha(1 + 2 \kappa) +\sqrt{\alpha^2 - 4\kappa (\kappa+1)}}{1+ \alpha^2}.
$$
This region exists only if if $\kappa\le (\sqrt{1 + \alpha^2}-1)/2.$ In this region the speed may be given by the SW or by the KPP value. We know with certainty that the speed will be given by the KPP value when the upper and lower bounds in (\ref{bounds}) coincide. The simplest condition to ensure this regime is to require $f''(u)<0$.  This condition is met provided $h_a> 4 R(h_a)$. One can show that $f''(u)<0$ for an applied field in the range
$$
\frac{ 4 ( \alpha(1 + 2 \kappa) -\sqrt{\alpha^2 - 64\kappa (\kappa+1)}\, )}{16+ \alpha^2} \le  \frac{h_a}{h_w} \le \frac{4 (  \alpha(1 + 2 \kappa) +\sqrt{\alpha^2 - 64\kappa (\kappa+1)}\, )}{16+ \alpha^2}.
$$
 If this condition is fulfilled the time evolution of a pertubation to the unstable state $u= 0 \, (\theta= \pi)$ evolves into a monotonic travelling domain of speed $c_{KPP}.$ This criterion is sufficient but not necessary, 
  the transition from a pushed to a pulled front may occur before the upper and lower bounds in  Eq. (\ref{bounds})  coincide. In this problem for which there is an exact solution we know that the transition will occur when $c_{\text{\tiny{SW}}}  = c_{\text{\tiny{KPP}}} .$   This implies that the speed of the moving front will be, in the original dimensional variables, 
\be \label{final}
v=  \left\{
\begin{array}{lll}  
v_{\text{\tiny{SW}}}  & {\rm if} \quad 0<H < H_{-} \\
v_{\text{\tiny{KPP}}}  & {\rm if} \quad H_{-} < H  < H_+ \\
v_{\text{ \tiny{SW}}}  & {\rm if} \quad H_+ < H < H_W, 
 \end{array} \right.
\ee
where the Walker field is $H_W = 2 \alpha K_d/ (\mu_0 M_s)$, the limiting fields $H_{\pm}$ are those given in (\ref{hlimits})  and
\be \label{SWspeed}
v_{\text{\tiny{SW}}} =  \frac{H_a}{\alpha} \frac{\mu_0 |\gamma| \sqrt{C_{ex}}}{ \sqrt{ 2 K_u + K_d ( 1- \sqrt{1 - (H/H_W)^2})}},
\ee
\be\label{KPPspeed}
v_{\text{\tiny{KPP}}} = \frac{ 2   |\gamma | \sqrt{C_{ex}}}{\alpha M_s} \sqrt{ \mu_0 M_s H_a - 2 K_u -    K_d ( 1- \sqrt{1 - (H/H_W)^2})}.
\ee
This transition for the speed occurs only if $\alpha^2 > 16\kappa(\kappa+1)$ or equivalently, if
\be\label{Kdmin}
K_d > \frac{4 K_u}{\sqrt{4 + \alpha^2}-2}.
\ee
There is an explicit analytic solution for the domain wall profile in the Schryer Walker regime, a closed form analytic solution for the KPP front does not exist in this problem. We know that  it shares the qualitative features of the Walker solution, that is,  it is a monotonically decaying front joining the stable and unstable equilibrium points. 
In Fig. 1 the shaded region shows, for fixed $\kappa$  the range of applied field as a function of $\alpha$ where the KPP regime exists.  The field is expressed in units of the Walker field.

\begin{figure}[H]
\centering
\includegraphics[width=0.5 \textwidth]{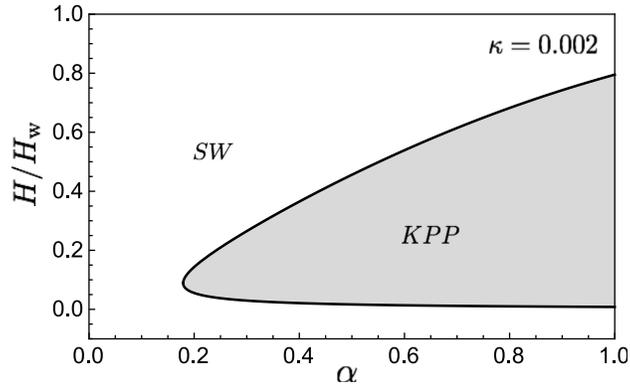}
\caption{Range of  applied field (in units of the Walker field) versus  $\alpha$ showing the region  for which the KPP regime exists for  $\kappa=0.002$ For low values of $\alpha$  this regime is not present and  the Schryer Walker profile is the selected solution. As $\kappa$ decreases the KPP regime extends to lower values of $\alpha$. }
\end{figure}

  The main effect of the KPP regime is to slow the rate of increase of the speed with the applied field. 
 In the figures below we show the speed and reaction functions $f(u)$ for different parameter values.  We fix the material parameters  $M_s=36000$ A/m, and
$C_{ex} = 10^{-13}$J/m.  The speed as function of magnetic induction $B = \mu_0 H$ for two different values of $\alpha$  with material parameters  $K_u = 40$ J m$^{-3},$   $K_d = 7500$ J m$^{-3},$ is shown in Fig. 2. The solid line is the speed given in Eq. (\ref{final}) and the dashed lines are the values of the SW speed in the region where the KPP regime holds. For larger values of $\alpha$ the rate of increase of the speed with the field is significantly slower than for the SW speed.

\begin{figure}[H]
\centering
\includegraphics[width=0.5 \textwidth]{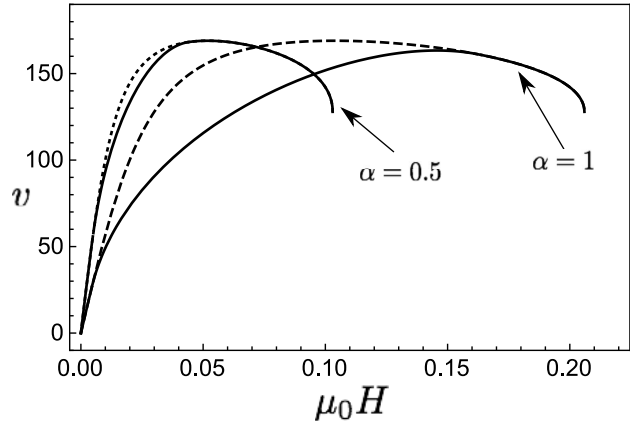}
\caption{The solid line is  the speed of the domain wall as a function of  $\mu_0 H$ for different values of $\alpha$. The dashed line is the Schryer-Walker speed in the region where the  speed is the KPP value. The difference between the SW speed and the KPP speed increases with $\alpha$.  }
\end{figure}

 The different speed regimes obey to the change in the reaction term as parameters are varied. In Fig. 3 we show the reaction term for $\alpha=1, K_u = 40$ J m$^{-3}$ and $K_d = 7500$ J m$^{-3},$ for different values of the field. When the applied field $B = \mu_0 H  =0.05$ the reaction term is of KPP type. In both cases the equilibrium $u=0$  ($\theta= \pi$) is unstable. 
 
\begin{figure}[H]
\centering
\includegraphics[width=0.5 \textwidth]{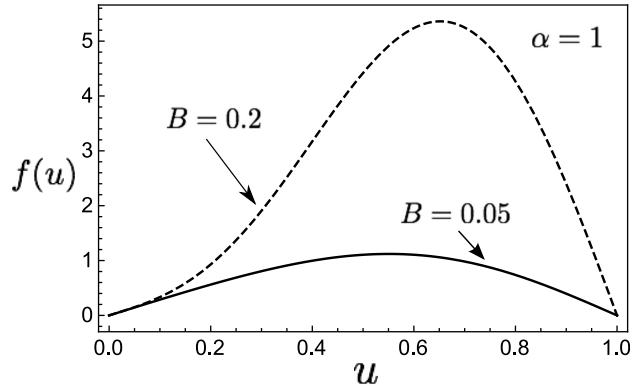}
\caption{Reaction term $f(u)$ for fixed $\alpha$ at two different values of the applied field.  The dashed line shows the reaction term for an applied field for which  the SW speed is selected, the solid line corresponds to a field for which the KPP speed is selected as shown in Fig. 2 }
\end{figure}

For smaller $\alpha$ with the same values for $K_u$ and $K_d$ the nature of the reaction terms changes drastically with the field. For $B=0.08$ the reaction term is bistable, the states $\theta=0 $ and $\theta = \pi$ are stable, and the KPP regime does not exist. There is a single traveling front, the SW solution.  For a small field the reaction term is of KPP type. Not only does the speed change but the stability of the equilibrium $\theta= \pi$ changes with the field. 

\begin{figure}[H]
\centering
\includegraphics[width=0.5 \textwidth]{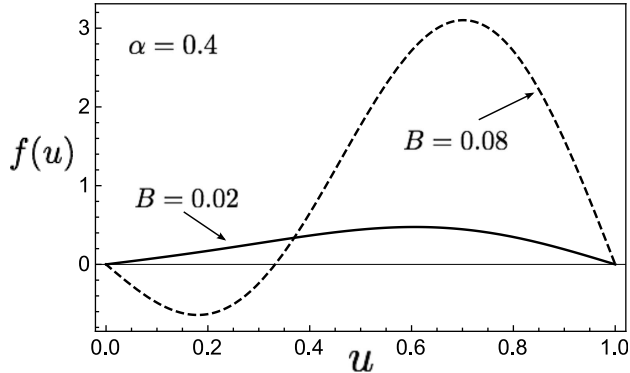}
\caption{
As in Fig. 3 for $\alpha=0.4$. Reaction term $f(u)$ for  $\alpha =0.4$ at two different values of the field. The dashed line depicts the reaction term for an applied field for which   the SW solution is  selected, the solid line corresponds to a field for which the KPP speed is selected.  In the bistable case ($B=0.08$) the SW domain wall solution is the unique front. }
\end{figure}

In Fig. 5 we show the speed as a function of field for different values of the hard axis anisotropy.  Here the presence of the KPP regime has a larger effect for larger values of $K_d$. 

\begin{figure}[H]
\centering
\includegraphics[width=0.5 \textwidth]{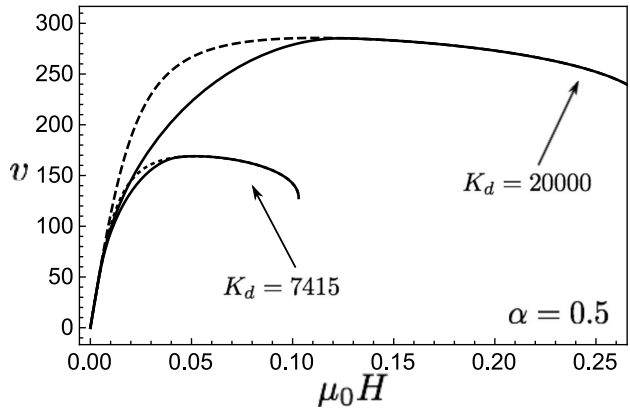}
\caption{As in Fig. 2, for different values of hard axis anisotropy.  }
\end{figure}

The existence of different regimes of front propagation in the reduced system studied by Schryer and Walker that we report here follows directly from the general theory of reaction diffusion equations. It is natural to ask whether this regime arises in the full LLG equations. There is analytical  evidence that it does in the case of very thin nanotubes as demonstrated in \cite{GoRoSl2014, De2014}. For  nanowires and thin films an asymptotic expansion of the LLG equation for large perpendicular anisotropy \cite{De2015}  shows the transition from the SW to the KPP solution at small field, transition which we identify with the transition point $H_{-}.$ In these works the assumption of fixed azimuthal angle is not imposed.  While, to our knowledge, the parameter ranges for which this transition occurs have not been accessed experimentally   they may become accesible in the future.  Not only that but effects which are neglected here, such as cubic anisotropy will likely modify the parameter range at which this behavior could be observed.  

 \begin{acknowledgments}
 This work has been partially supported by Fondecyt (Chile) projects 1141155, 1120836 and 
 Iniciativa Cient\'\i fica Milenio, ICM (Chile), through the Millenium Nucleus RC120002.
\end{acknowledgments}

\end{document}